\documentclass[aps,prd,twocolumn,showpacs,groupedaddress]{revtex4-1}

\usepackage{graphicx}
\usepackage{dcolumn}
\usepackage{bm}
\usepackage{amsmath}
\usepackage{amssymb}

\begin{document}

\title{Chiral Symmetry Breaking in Planar QED in External Magnetic Fields }

\author{Paolo Cea}
\affiliation{Dipartimento di Fisica dell'Universit\`a di Bari, I-70126 Bari, Italy \\
and INFN, Sezione di Bari, I-70126 Bari, Italy}
\email{paolo.cea@ba.infn.it}

\author{Leonardo Cosmai}
\affiliation{INFN, Sezione di Bari, I-70126 Bari, Italy}
\email{leonardo.cosmai@ba.infn.it}

\author{Pietro Giudice}
\affiliation{Department of Physics, College of Science, Swansea University, Singleton Park, SA2 8PP Swansea, United Kingdom}
\email{p.giudice@swansea.ac.uk}

\author{Alessandro Papa}
\affiliation{Dipartimento di Fisica dell'Universit\`a della Calabria,
I-87036 Rende (Cosenza), Italy \\
and INFN, Gruppo collegato di Cosenza, I-87036 Rende (Cosenza), Italy}
\email{alessandro.papa@cs.infn.it}

\date{\today}

\begin{abstract}
We investigate planar quantum electrodynamics (QED) with two degenerate staggered fermions in an external magnetic field on the lattice. We argue  that in  external magnetic fields there is  dynamical generation of mass for two-dimensional massless Dirac fermions in the weak-coupling region. We extrapolate our lattice results to the quantum Hall effect in  graphene.
\end{abstract}


\maketitle

\section{Introduction}
\label{sec1}
Quantum electrodynamics (QED) in 2+1 dimensions is interesting as a model for several condensed matter systems. In fact, quantum electrodynamics with two massless Dirac fermions could be relevant to describe the low-energy excitations of a single sheet of carbon atoms arranged in a honeycomb structure called 
``graphene''~\cite{Novoselov:2004aa,Novoselov:2005aa}. When graphene is immersed in a transverse magnetic field, the presence of Landau levels at zero energy leads
to half-integer quantum Hall effect. Moreover, for very strong magnetic fields there is experimental evidence for the dynamical generation of a gap, which
 signals the spontaneous breaking of the chiral symmetry. In fact, it has been suggested that a magnetic field is a strong catalyst of chiral symmetry breaking in spinorial QED~\cite{Gusynin:1994re,Gusynin:1994va} even at the weakest attractive interaction between fermions. \\
The aim of the present paper is to investigate, by means of non-perturbative Monte Carlo simulations, planar quantum electrodynamics (QED) with two degenerate staggered fermions in an external magnetic field. To make contact with the physical planar systems, we choose to work in the  weak-coupling region.  A preliminary
account of the results discussed in the present paper has been published in Ref.~\cite{Cea:2011hu}. \\
The plan of the paper is as follows. In Sect.~\ref{sec2}, for completeness,
we briefly discuss our method to introduce  background fields on the lattice and compare
with different approaches in the literature. Section~\ref{sec3}  is devoted to the discussion of our lattice  Euclidean action. In Sect.~\ref{sec4} we present
the results of our  numerical simulations for two different values of the gauge coupling in the weak-coupling region.  In Sect.~\ref{sec5} we extrapolate our results to
the physical relevant case of the quantum Hall effect in graphene. Finally, our conclusions are relegated in  Sect.~\ref{sec6}.
\section{Background fields on the lattice}
\label{sec2}
The study of lattice gauge theories with an external background field has been pioneered in Ref.~\cite{Damgaard:1987ec,Damgaard:1988hh} for the U(1) Higgs model
in an external electromagnetic field. In the continuum a background field can be introduced by writing:
\begin{equation}
\label{eq2.1}
A_{\mu}(x) \;  \rightarrow  \; A_{\mu}(x)  \; + \; A^{\text{ext}}_{\mu}(x)  \; .
\end{equation}
In the lattice approach one deals with link variables $U_{\mu}(x)$. Accordingly, on the lattice Eq.~(\ref{eq2.1}) becomes:
\begin{equation}
\label{eq2.2}
U_{\mu}(x) \;  \rightarrow  \; U_{\mu}(x) \, U^{\text{ext}}_{\mu}(x)  \; ,
\end{equation}
where $U^{\text{ext}}_{\mu}(x)$ is the lattice version of the background field $A^{\text{ext}}_{\mu}(x)$. As a consequence the
gauge action gets modified as:
\begin{equation}
\label{eq2.3}
S_{G}[U] \;  \rightarrow  \;S_{G}[U]   \; +  \delta \, S[U,U^{\text{ext}}] \; ,
\end{equation}
where $ \delta \, S[U,U^{\text{ext}}]$ takes into account the influence of the external field~\cite{Smit:1986fn,Ambjorn:1989qr,Ambjorn:1990wf,Kajantie:1998rz,Buividovich:2009wi,D'Elia:2010nq,Bali:2011qj,Ilgenfritz:2012fw}. An alternative method, which is equivalent in the continuum limit,
 is based on the observation that an external background field can be introduced via an external 
 current~\cite{Cea:1989ag,Cea:1990td,Levi:1995kc,Ogilvie:1997my,Chernodub:2001da}:
\begin{equation}
\label{eq2.4}
J^{\text{ext}}_{\mu}  \;  = \; \partial_{\nu} \;   F^{\text{ext}}_{\nu \mu}  \; .
\end{equation}
The gauge action gets modified in an obvious manner:
\begin{equation}
\label{eq2.5}
S_{G}  \;  \rightarrow  \;S_{G}  \; +  \; S_{B} \; , 
\end{equation}
where:
\begin{eqnarray}
\label{eq2.6}
S_{B}  \; &= \;  \int dx  \; J^{\text{ext}}_{\mu} (x)  \; A_{\mu}(x)  \nonumber \\ \; &= \; - \; \frac{1}{2} \;  \int dx  \;  F^{\text{ext}}_{\nu \mu}(x) \,  F_{\nu \mu}(x) \;.
\end{eqnarray}
The background action $S_{B}$  can be now easily discretized on the lattice. \\
The main disadvantage of this approach resides on the fact that it cannot be extended to the case of non-Abelian gauge group
in a gauge-invariant way. To overcome this problem, the background field on the lattice can be implemented by means of 
the gauge invariant lattice Schr\"odinger functional~\cite{Cea:1996ff,Cea:1999gn}:
\begin{equation}
\label{eq2.7}
 {\mathcal{Z}}[U^{\mathrm{ext}}_k] = \int {\mathcal{D}}U \; e^{-S_G}  \; ,
\end{equation}
where the functional integration is extended over links on a lattice with the
hypertorus geometry  and satisfying the constraints ($x_t$ is the  temporal coordinate)
\begin{equation}
\label{eq2.8}
U_k(x)|_{x_t=0} = U^{\mathrm{ext}}_k(\vec{x}) \; .
\end{equation}
We also impose that links at the spatial boundaries are fixed according to Eq.~(\ref{eq2.8}). In the continuum this last
condition amounts to the requirement that fluctuations over the background field vanish at infinity. \\
The effects of dynamical fermions can be accounted for quite easily. In fact, when including dynamical fermions, the lattice Schr\"odinger functional
in presence of a static external background gauge field becomes~\cite{Cea:2004ux}
\begin{eqnarray}
\label{eq2.9}
\mathcal{Z}[U^{\mathrm{ext}}_k]  &=& 
\int_{U_k(L_t,\vec{x})=U_k(0,\vec{x})=U^{\text{ext}}_k(\vec{x})}
\mathcal{D}U \,  {\mathcal{D}} \psi  \, {\mathcal{D}} \bar{\psi} e^{-(S_G+S_F)} 
\nonumber \\ 
&=&  \int_{U_k(L_t,\vec{x})=U_k(0,\vec{x})=U^{\text{ext}}_k(\vec{x})}
\mathcal{D}U e^{-S_G} \, \det M \,,
\end{eqnarray}
where $S_F$ is the fermionic action and $M$ is  the fermionic matrix.
Notice that the fermionic fields are not constrained and
the integration constraint is only relative to the gauge fields.
This leads  to the appearance of  the gauge invariant fermionic determinant after integration on the 
fermionic fields.  As usual we impose on fermionic fields
periodic boundary conditions in the spatial directions and
antiperiodic boundary conditions in the temporal direction.
\section{Lattice planar QED in external magnetic field }
\label{sec3}
We are interested in planar quantum electrodynamics  with $N_f=2$ degenerate Dirac fields in
an external constant magnetic field. As it is well known,  Dirac fields  are described 
non-perturbatively by the lattice Euclidean action using $N$ flavours of
staggered fermion fields $\overline\chi,\chi$~\cite{Hands:2002dv}:
\begin{equation}
\label{eq3.1}
 S=S_G+\sum_{i=1}^{N} \sum_{n,m}  \overline\chi_i(n)
M_{n,m} \chi_i(m)\;,
\end{equation}
where $S_G$ is the gauge field action and the fermion matrix is given by:
\begin{eqnarray}
\label{eq3.2}
&& M_{n,m}[U] = \nonumber \\ && \sum_{\nu=1,2,3} \frac{\eta_{\nu}(n)}{2} \left\{ U_{\nu}(n)
      \delta_{m,n+\hat{\nu}}- 
     U_{\nu}^{\dagger}(m)  \delta_{m,n-\hat{\nu}} \right\}   + m_0  \delta_{m,n} \; ,\nonumber  \\    
&&   \; \; \; \;  \eta_{\nu}(n) = (-1)^{n_1+\ldots+n_{\nu-1}} \; ,
\end{eqnarray}
where $m_0$ is the bare fermion mass.
Here we adopt the  compact formulation for the electromagnetic field (for a detailed account see Ref.~\cite{Fiore:2005ps}). 
The gauge action is:
\begin{equation}
\label{eq3.3}
S_G[U]= \beta \sum_{n,\mu < \nu}{ \left[ 1 - \frac{1}{2} \left( U_{\mu \nu}
 (n)+ U_{\mu \nu}^{\dagger}(n) \right) \right]}\;,
\end{equation}
where $U_{\mu \nu}(n)$ is the plaquette and $\beta=\frac{1}{e^2 }$.  The action Eq.~(\ref{eq3.1}) with $N=1$ flavours of
staggered fermions corresponds  to $N_f=2$ flavours of 4-component Dirac fermions $\Psi$~\cite{Burden:1986by}. \\
To introduce an external magnetic field, we shall follow the  lattice Schr\"odinger functional described 
in Sect.~\ref{sec3} (for a different  approach see Ref.~\cite{Alexandre:2001pa}). 
Accordingly, in the functional integration over the lattice links we constrain
the spatial links belonging to the  time slice  $x_t=0$  to
\begin{equation}
\label{eq3.4}
U_k(\vec{x},x_t=0) = U^{\text{ext}}_k(\vec{x})
\,,\,\,\,\,\, k=1,2 \,\,,
\end{equation}
 $U^{\text{ext}}_k(\vec{x})$ being the lattice version  of the external continuum
gauge potential. Since our  background field does not vanish at infinity,
 we
must also impose that, for each time slice $x_t \ne 0$, spatial links exiting
from sites belonging to the spatial boundaries  are fixed according to Eq.~(\ref{eq3.4}).  \\
The continuum  gauge potential giving rise to a constant magnetic field is given by:
\begin{equation}
\label{eq3.5}
A^{\text{ext}}_k(\vec{x}) =  \delta_{k,2} \ x_1 H \; ,
\end{equation}
so that:
\begin{equation}
\label{eq3.6}
U^{\text{ext}}_1(\vec{x}) = 1 \, , 
\,  U^{\text{ext}}_2(\vec{x}) = \cos(  eHx_1) +
i \sin( eHx_1 )   \,.
\end{equation}
Since our lattice has the topology of a torus, the magnetic field turns out to be quantized:
\begin{equation}
\label{eq3.7}
  e H = \frac{2 \pi}{L}
n_{\text{ext}} \; , \;\;\;  n_{\text{ext}}\,\,\,{\text{integer}} \; ,
\end{equation}
where $L$ is the lattice size. We recall once more that the fermion fields are unconstrained and satisfy antiperiodic boundary conditions in the timelike 
direction and  periodic boundary conditions in the spatial directions. \\
Our numerical results were obtained by simulating the action Eq.~(\ref{eq3.1}) on $L^3$ lattice using standard hybrid Monte Carlo
algorithm.  
\section{Chiral symmetry breaking}
\label{sec4}
\begin{figure*}
\includegraphics[width=0.9\textwidth,clip]{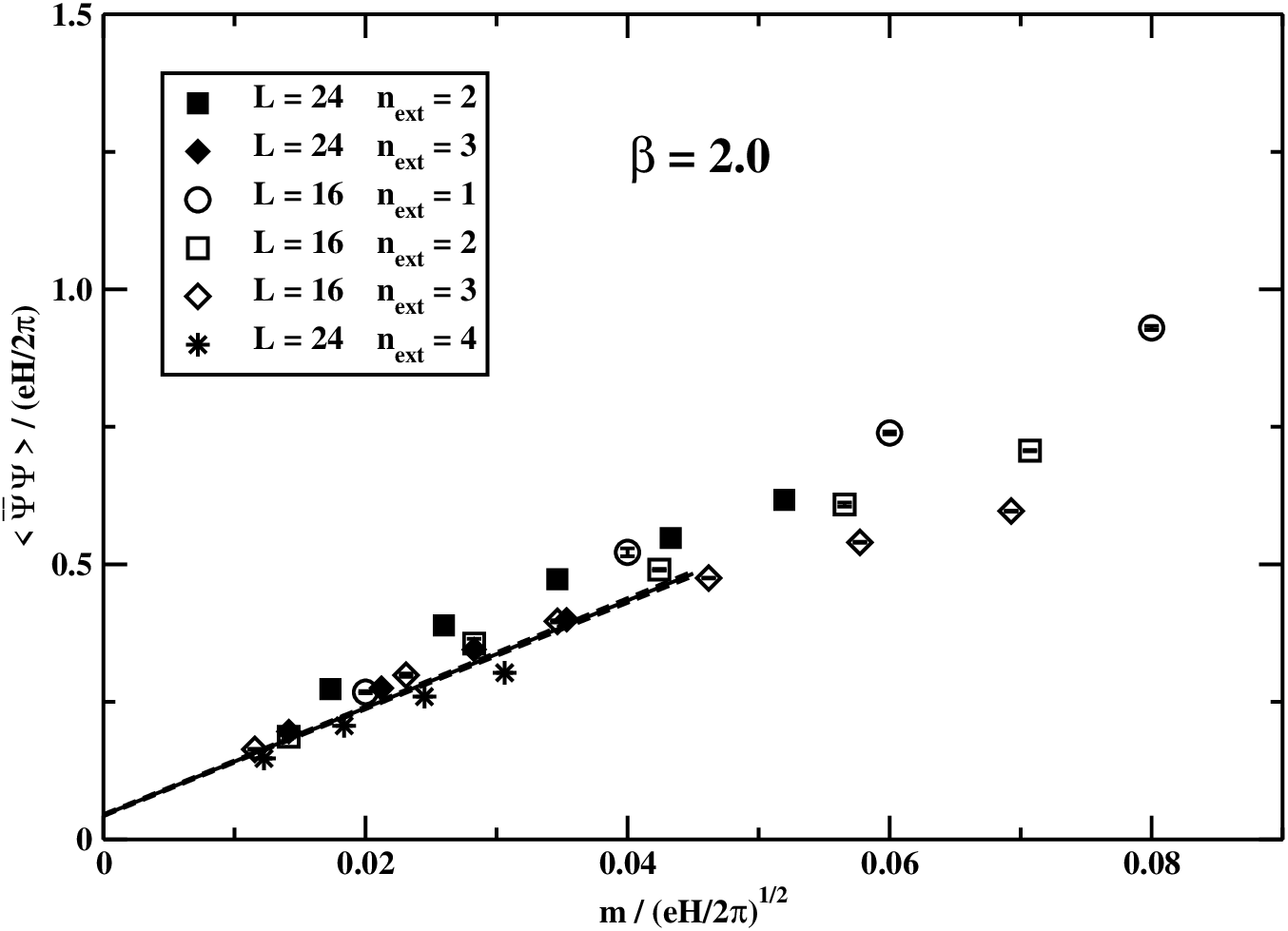}
\caption{\label{Fig-1}
Scaled chiral condensate versus the scaling variable $x = \frac{m_0}{\sqrt{eH/ 2 \pi}}$ for $\beta = 2.0$. The continuum line is the linear fit of the data 
Eqs.~(\ref{eq4.2}) and (\ref{eq4.3}) in the scaling region $0 < x \lesssim 0.045$.}
\end{figure*}
\begin{figure*}
\includegraphics[width=0.9\textwidth,clip]{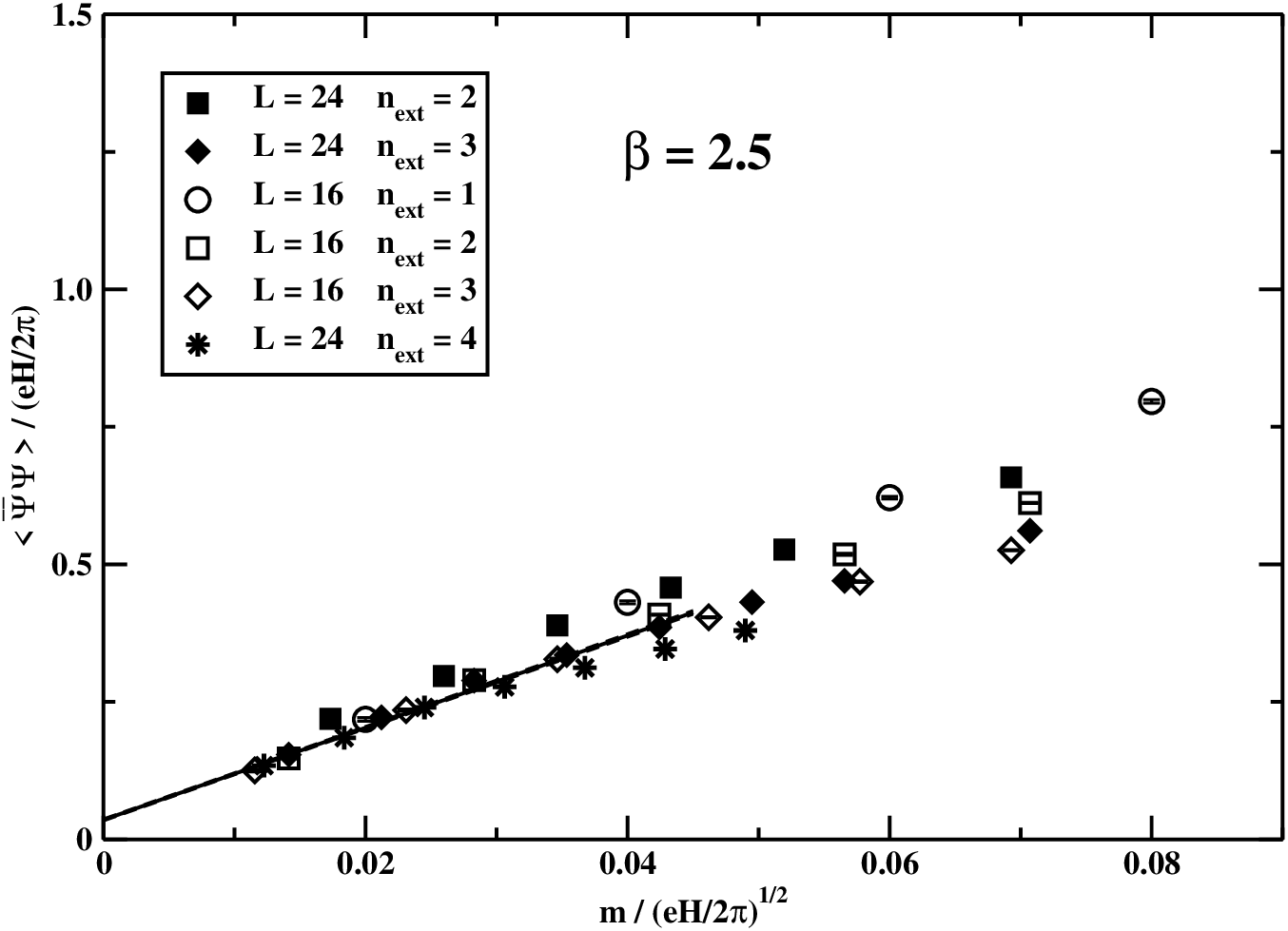}
\caption{\label{Fig-2}
Scaled chiral condensate versus the scaling variable $x = \frac{m_0}{\sqrt{eH/ 2 \pi}}$ for $\beta = 2.5$. The continuum line is the linear fit of the data 
Eqs.~(\ref{eq4.2}) and (\ref{eq4.3}) in  the scaling region $0 < x \lesssim 0.045$.}
\end{figure*}

 We are looking for the dynamical generation of a gap for massless fermions. This corresponds to
a non-zero chiral condensate $ \langle\overline\Psi\Psi\rangle$ in the chiral limit.  
Our strategy is to measure the fermion condensate with a  small bare fermion mass $m_0$ and then perform
the massless limit $m_0 \rightarrow 0$ in presence of a constant external magnetic field. Our simulations
have been performed in the weak-coupling region with two different values of the gauge coupling
$\beta=2.0$ and $\beta=2.5$.  In fact, in the   weak-coupling region we expect that the effects of the Coulomb interactions
could be neglected allowing to extrapolate our numerical results to physical planar systems. \\
We have performed simulations on lattices with $L=16,$ 24 and $0.005 \leqslant m_0 \leqslant 0.03$ with
different strengths of the external magnetic field labelled by the integer $n_{\text{ext}}$ according to Eq.~(\ref{eq3.7}). 
 For each parameter set, to allow thermalization we  discard  $10000$ sweeps for $L=16$ and  $7000$ sweeps for $L=24$.
We collect about $50000$ hybrid Monte Carlo trajectories.  To optimize the performance of the hybrid Monte Carlo
algorithm, we tuned the simulation parameters to give an acceptance of about $80 \%$.
The chiral condensate  $ \langle\overline\Psi\Psi\rangle$ was estimated by the  stochastic source method.
 In order to reduce autocorrelation effects, measurements were taken every 10 steps for $L=16$  and every 5 steps for $L=24$ . 
 Data were analyzed by   the jackknife method combined with binning.  \\
In Fig.~\ref{Fig-1} we display the chiral condensate for different values of the lattice size, bare fermion mass, and magnetic field strength for $\beta=2.0$ . 
Note that, according to  Eq.~(\ref{eq3.7}), the strength of the external magnetic field depends on $n_{\text{ext}}$ as well on the lattice size $L$.
To avoid lattice discretization and finite volume effects, we have fixed the magnetic field  strength such that the magnetic length satisfies the bounds:
\begin{equation}
\label{eq4.1}
1 \ll \sqrt{{\frac{2 \pi}{eH}}} \; \ll  \; L \; .
\end{equation}
We expect that in the continuum limit the relevant scale is set by the magnetic length. This means that the rescaled chiral condensate 
$\frac{\langle\overline\Psi\Psi\rangle}{\frac{eH}{2 \pi}}$  would depend only on the scaling variable $x \equiv \frac{m_0}{\sqrt{\frac{eH}{2 \pi}}}$. 
Actually, from  Fig.~\ref{Fig-1}, where we display the rescaled chiral condensate versus the dimensionless scaling variable $x$,
we see that in the region $x  \gtrsim  0.05$  data are rather scattered. However, in
the region $ x \lesssim 0.05$  our data seem to collapse to an universal curve. This means that in this region, that we shall call
the {\it scaling} region, the rescaled chiral condensate depends only on the scaling variable $x$. 
This allows us to extract the chiral condensate in the chiral limit $m_0 \rightarrow 0$, which corresponds to $x \rightarrow 0$,
for a fixed strength on the external magnetic field. In fact, we try to fit the data in the scaling region  $0 < x \lesssim 0.045$
according to:
\begin{equation}
\label{eq4.2}
 \frac{\langle\overline\Psi\Psi\rangle}{\frac{eH}{2 \pi}} \; = \; a_0 \, + \, a_1 \, x \; \; , \;\;\;\; \; x = \frac{m_0}{\sqrt{\frac{eH}{ 2 \pi}}} \, .
\end{equation}
The best fit of the data to Eq.~(\ref{eq4.2}) in the scaling region gives:
\begin{eqnarray}
\label{eq4.3}
&& a_0 \, = \, 0.04399 \, \pm \, 0.00131 \;  \; , \; \;\;\;\;
  a_1 \, = \, 9.759 \, \pm \, 0.055 \;  \; , \; \;\;\;\; \nonumber \\
&&  \chi^2_{\rm d.o.f.} \; \simeq 747 \; .
\end{eqnarray}
We note, however, that there are sizable violations of our  scaling  law as implied by the huge reduced chi-square. We believe that
these scaling violations are mainly due to the fermion  interactions with the electromagnetic field, which  could 
introduce a spurious dependence of the  scaled chiral condensate on the dimensionless ratio $ \frac{e^2}{\sqrt{\frac{eH}{2 \pi}}}$. 
To check this point, we have performed  numerical
simulations by increasing the gauge coupling $\beta$ (which corresponds to a smaller $e^2$). In fact, in Fig.~\ref{Fig-2} we display 
the results of our simulations  for $\beta=2.5$.  Again we see that the data for the rescaled chiral condensate  seem to collapse to  
 to an universal curve in the scaling region $ x \lesssim 0.05$. Moreover,  comparing  Fig.~\ref{Fig-2} with Fig.~\ref{Fig-1} it is evident that
the scaling violation are greatly reduced allowing a better extrapolation to the chiral limit. Fitting the data to  Eq.~(\ref{eq4.2})
we find: 
\begin{eqnarray}
\label{eq4.4}
&& a_0 \, = \, 0.03544 \, \pm \, 0.00084 \;  \; , \; \;\;\;\;
a_1 \, = \, 8.382 \, \pm \, 0.032\;  \; , \;\;\;\;\;  \nonumber \\ &&\chi^2_{\rm d.o.f.} \; \simeq 466 \; .
\end{eqnarray}
Even though the reduced chi-square is quite large, we believe that our results are robust enough to allow the extrapolation
of the chiral condensate to the chiral limit.
As a consequence, we conclude that in the chiral limit the external magnetic field does induce
a non-zero chiral condensate. From Eqs.~(\ref{eq4.2}) and (\ref{eq4.4}) we find for the chiral condensate
in the massless limit:
\begin{equation}
\label{eq4.5}
\langle\overline\Psi\Psi\rangle  \; = \;  \frac{ eH}{2 \pi}  \left(  0.03544 \, \pm \, 0.00084 \right) \, .
\end{equation}
The non-zero value of the chiral condensate can be interpreted as the generation of a dynamical fermion
mass which, in principle, can be extracted from the  non-zero chiral condensate in the chiral limit. \\
In the determination of the value of the chiral condensate, as given
in Eq.~(\ref{eq4.5}), we neglected a possible contribution present 
even in absence of the magnetic field. Indeed, in 
Ref.~\cite{Armour:2011zx} it was shown that the chiral condensate is non-zero 
in the weak-coupling regime of compact planar QED even at zero external
magnetic field. In order to check the possible impact of this zero-field
contribution on our determination of the chiral condensate, we observe that 
in Ref.~\cite{Fiore:2005ps} two of us found 
$\beta^2 \langle\overline\Psi\Psi\rangle  \; \approx \; 1.5 \times 10^{-3}$
for $H=0$ on a lattice with $L=12$. This 
result implies, for $\beta=2.5$, that $\langle\overline\Psi\Psi\rangle  \; = \; 0.00024$, in 
lattice units. In this work the smallest value of the chiral condensate 
induced by an external magnetic field is obtained for $n_{\text{ext}}=1$ and 
$L=16$, which implies $eH/(2\pi)=0.0625$ and therefore, through 
Eq.~(\ref{eq4.5}), $\langle\overline\Psi\Psi\rangle \;=\; 0.002215$. 
The latter value is one order of magnitude bigger than the former and
cannot be attributed to finite size effects, in consideration of the
similar lattice sizes adopted in the two determinations. This allows us
to safely neglect the zero-field contribution to the chiral condensate.

\section{Extrapolation to Graphene}
\label{sec5}
In this Section we attempt to apply our numerical determination of the chiral condensate in the chiral limit to graphene immersed in a transverse
magnetic field. For the reader's convenience, we briefly discuss the remarkable quantum Hall effect in graphene. \\
As is well known, graphene is a flat monolayer of carbon atoms tightly packed in   a two dimensional honeycomb lattice consisting of two interpenetrating triangular sublattices (for a review, see  Ref.~\cite{CastroNeto:2009zz}).
Indeed, the structure of graphene has attracted considerable attention since  the low-energy excitations  are given by two Pauli spinors $\Psi_{\pm}$  which satisfy the massless two-dimensional Dirac equation with the speed of light replaced by the Fermi velocity $v_F  \simeq  1.0 \, 10^8  \,$ cm/s. The Pauli spinors can be combined into a single Dirac spinor $ \Psi=\left(\begin{array}{c}\Psi_+  \\ \Psi_- \end{array} \right)$. Taking into account the real spin degeneracy, we see
 that the low-energy dynamics of graphene can be accounted for by $N_f=2$ massless  Dirac fields~\cite{Drut:2009aa,Drut:2009aj}. \\
\begin{figure*}
\includegraphics[width=0.8\textwidth,clip]{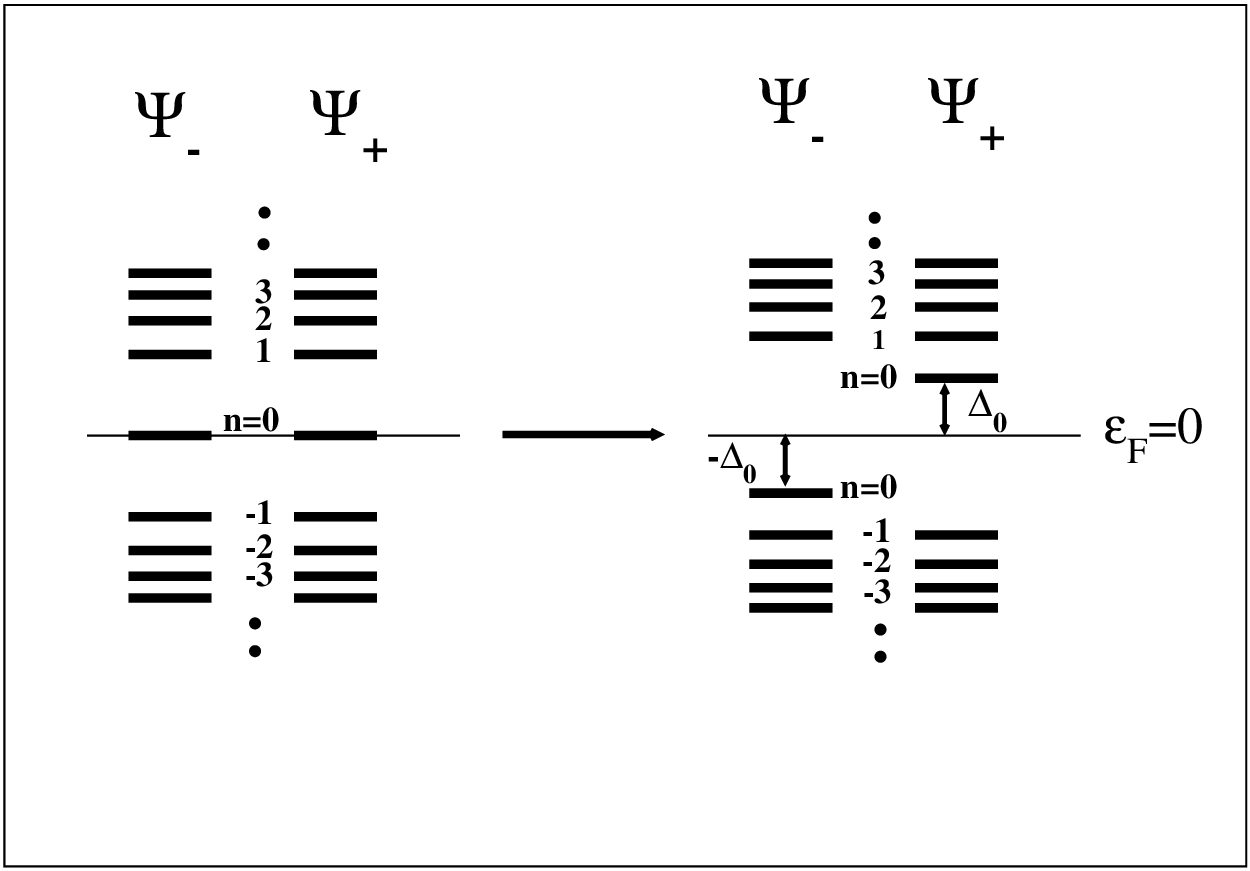}
\caption{\label{Fig-3}
Schematic spectrum of Landau levels of graphene in applied magnetic field (left). Landau levels with  dynamical generation of a gap $\Delta_0$ (right).  
The Fermi level is at $\varepsilon_F =  0$.}
\end{figure*}
When graphene is immersed in a  transverse magnetic field, the relativistic massless dispersion of the electronic wave functions results in non-equidistant Landau levels 
\footnote{ In this Section we use cgs units.}:
\begin{equation}
\label{eq5.1}
\varepsilon_n =  {\rm sign}(n) \sqrt{2 |n| \hslash \frac{ v_F^2}{c} eH} \, , \,  \;\;\; n \, = \, 0 \, , \, \pm \, 1 \, , \, \pm \, 2  \; ...
\end{equation}
where $eH>0$, $e$ being the elementary charge (see Fig.~\ref{Fig-3}, left). The presence of anomalous Landau levels at zero energy, $\varepsilon_0=0$, leads to half-integer quantum Hall  effect corresponding  to quantized filling factor $\nu   =  \pm 2 \, , \,  \pm 6 \, , \, \pm 10  \,\,   ...$. \\
Recent  studies of quantum Hall effect in graphene in very strong magnetic field $H \gtrsim 20 \, T$ (1 T = $10^4 \, $ gauss) have revealed new quantum Hall states corresponding to filling factor $\nu   = 0 \, , \, \pm  1 \, , \,  \pm 4$~\cite{Zhang:2006aa,Jiang:2007aa}. The new plateaus at $\nu   = 0 \,  \, , \,  \pm 4$ can be explained by Zeeman spin splitting. On the other hand the  $\nu   =  \pm 1$ plateaus are associated with the spontaneous breaking of the symmetry in the $n=0$ Landau levels
(the so-called {\it valley} symmetry). Indeed, these states are naturally explained if there is dynamical generation of a gap $\Delta_0$ (see Fig.~\ref{Fig-3}, right).  \\
The gap $\Delta_0$ can be extracted from the measured activation energy. In fact,  in Fig.~\ref{Fig-4} we display the  measured activation energy gap 
$\Delta E (\nu = 1)$ as a function of the magnetic field for the $\nu = 1$ quantum Hall states~\cite{Jiang:2007aa}.  
To extract  $\Delta_0$ from the activation energy data, we need to take care of the Zeeman energy which for strong magnetic fields is no more negligible. 
To this end, we may fit the data to:
\begin{figure*}
\includegraphics[width=0.9\textwidth,clip]{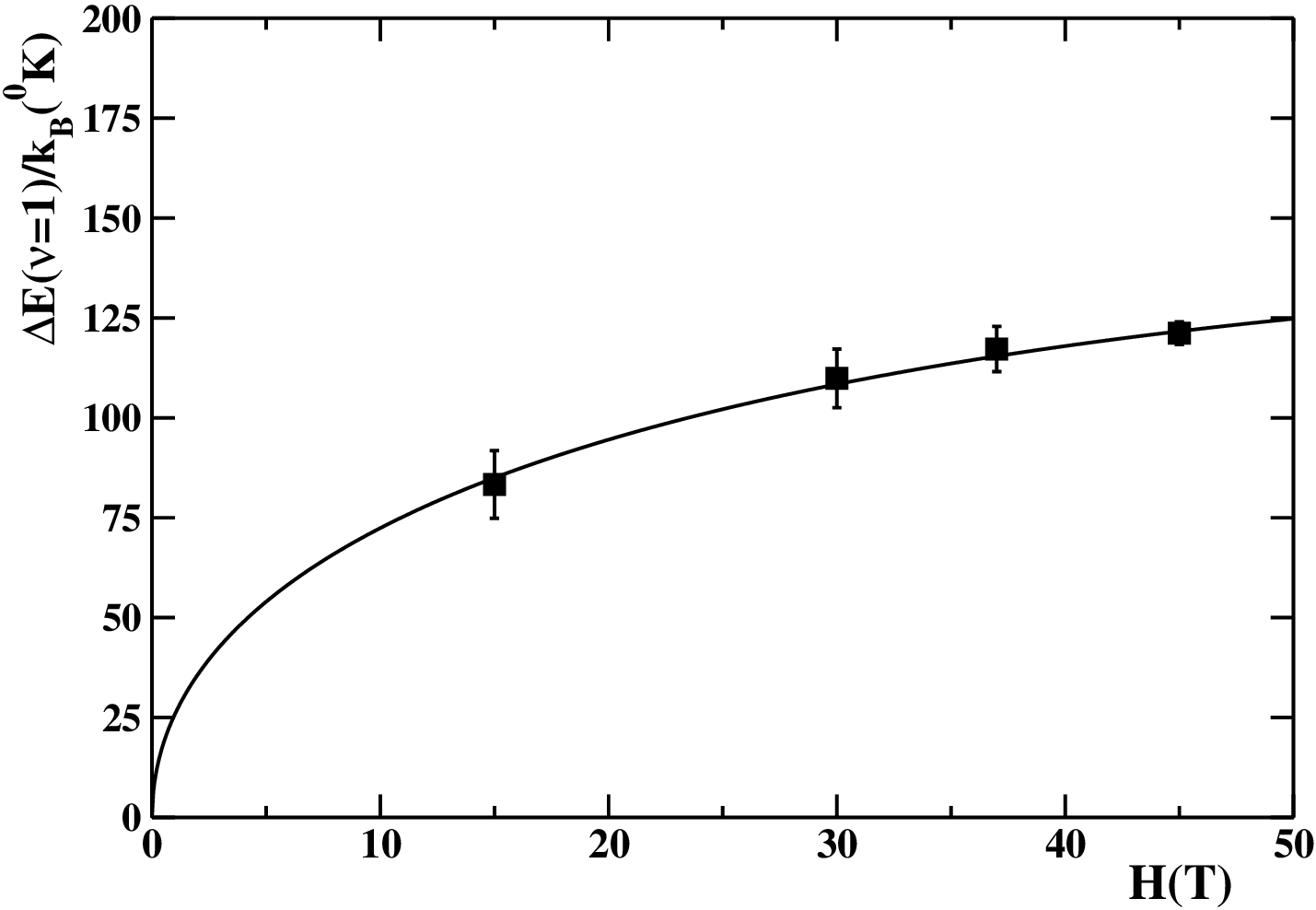}
\caption{\label{Fig-4}
The measured activation energy gap $\Delta E (\nu = 1)$ as a function of magnetic field for the quantum Hall states at filling factor $\nu = 1$. 
The data have been extracted from Fig.~2 of Ref.~\cite{Jiang:2007aa}. The continuum line is the best fit of the experimental data to Eq.~(\ref{eq5.2}).}
\end{figure*}
\begin{equation}
\label{eq5.2}
\Delta E (\nu = 1)  =  2 \, \left ( \Delta_0(H) \, + \,  \frac{g}{2} \mu_B \, H \right ) \; ,
\end{equation}
where $\mu_B$ is the Bohr magneton,  $g=2$ and $\Delta_0(H) \sim \sqrt{H}$~\cite{Jiang:2007aa}.  Figure~\ref{Fig-4}  shows that, indeed, our
Eq.~(\ref{eq5.2}) gives an excellent fit to the data.  We find:
\begin{equation}
\label{eq5.3}
\Delta_0(H) \, = \, (13.57 \, \pm 0.28) \; {\rm K} \, k_B \; \sqrt{H(T)}
\; ,
\end{equation}
where $H(T)$ means that the magnetic field is measured in Tesla. \\
Our strategy is, now, to relate the gap $\Delta_0$ to the chiral condensate. After that, using our determination of the chiral condensate on the lattice,
we will estimate the gap and compare with the experimental determination Eq.~(\ref{eq5.3}). \\
To this purpose we follow Ref.~\cite{Cea:2011vi}, where the hypothesis of 
rearrangement of the Dirac sea of graphene in an external magnetic field 
was used and the electron-electron Coulomb interactions were neglected.
Note that in graphene the electron-electron Coulomb interaction, $e^2/r$, in 
general is not small, so that this approximation could be questionable.
A direct calculation gives~\cite{Cea:2011vi}:
\begin{eqnarray}
\label{eq5.4}
&& \langle \overline\Psi \Psi \rangle \; = \; - \, 2 \,\Delta_0 \, \frac { \hslash  c  eH} {2 \pi}  \frac{1}{\sqrt{2  \hslash \frac{v_F^2}{c} eH}}
\frac{\Gamma(\frac{1}{2})} {\sqrt{ \pi}} \;  \zeta(\frac{1}{2}, 1+\alpha^2) \; , 
\nonumber \\
&& \; \; \; \; \; \; \;  \alpha \; = \;   \frac{\Delta_0}{\sqrt{2  \hslash  \frac{v_F^2}{c} eH}} \;  \; ,
\end{eqnarray} 
where  $\Gamma(x)$ is the Euler gamma function and  $\zeta(x,y)$ is the generalized Riemann Zeta function.
For small gap, we may  expand to the first order in $\Delta_0$. Using $\Gamma(\frac{1}{2}) = \sqrt{ \pi}$ and $\zeta(x,1) =  \zeta(x)$, we get:
\begin{equation}
\label{eq5.5}
\frac{\langle \overline\Psi \Psi \rangle}{ \frac { \hslash  c  eH} {2 \pi} } \; \simeq \; - \, 2 \,\Delta_0 \; \frac{1}{\sqrt{2  \hslash \frac{v_F^2}{c}  eH}}
 \;  \zeta(\frac{1}{2}) \; .
\end{equation}
This last equation relates the gap $\Delta_0$ to the rescaled dimensionless chiral condensate. Using our determination on the lattice for the
rescaled chiral condensate, we obtain:
\begin{equation}
\label{eq5.6}
\Delta_0  \; \simeq \;   \, - \,  \frac{\sqrt{\pi}}{  \zeta(\frac{1}{2})}  \; \;  \sqrt{ \frac{\hslash  \frac{v_F^2}{c} eH}{2\pi}}  \; \;  a_0  \; ,
\end{equation}
where $a_0$ is given in Eq.~(\ref{eq4.4}).  Finally, with the experimental value for the Fermi velocity we get:
\begin{equation}
\label{eq5.7}
\Delta_0(H) \, \simeq \, 2.6  \; \; {\rm K} \, k_B \; \sqrt{H(T)} .
\end{equation}
Comparing Eq.~(\ref{eq5.7}) with Eq.~(\ref{eq5.3}),  we see that our estimate of the gap is about a factor five smaller than the experimental data. However,
it is remarkable  that  we are able to reproduce the dependence on the external magnetic field $\Delta_0 \sim \sqrt{H}$. 
\section{Conclusions}
\label{sec6}
We investigated planar quantum electrodynamics  with two degenerate staggered fermions in an external magnetic field on the lattice. 
Our numerical results seem to indicate that  in an external magnetic field there is a non-zero
chiral condensate in the chiral limit pointing to a dynamical generation of mass for two-dimensional massless Dirac fermions.

We performed our simulations in the weak-coupling regime of the  compact formulation of the lattice gauge action. As discussed in 
Ref.~\cite{Armour:2011zx}, the {\it non-compact} formulation of the theory could have a different continuum limit than the compact one, the 
signature of this being the different magnetic monopole dynamics, which in compact QED leads to an enhanced chiral condensate.
As a matter of fact, in the compact theory the chiral condensate is non-zero in the strong-coupling regime and undergoes a crossover
to a non-zero value in the weak-coupling regime, while in the weak-coupling regime of the non-compact theory it is compatible with zero. Although we believe
that the numerical impact of this possible different behaviour in the continuum should be negligible to our purposes, we plan to explicitly check
this point by performing numerical simulations with the non-compact lattice gauge action. This will also permit us to make a comparison with the results 
of Ref.~\cite{Alexandre:2001pa}, where a different approach was adopted to introduce the background magnetic field on the lattice. 

We also tried to extrapolate our lattice results to the quantum Hall effect in graphene, since the low energy dynamics of graphene is described by $N_f=2$ 
massless Dirac fermions. Our non-perturbative Monte Carlo simulations allowed to confirm the dynamical breaking of the valley symmetry in the lowest Landau levels. 
Moreover, even though we greatly underestimate the dynamical gap, we were able to reproduce the dependence of the dynamical gap on the strength of the external 
magnetic field.

%

\end{document}